# Two-dimensional Model of a High-$T_c$ Superconducting Dynamo

Leonid Prigozhin and Vladimir Sokolovsky

*Abstract*—High temperature superconducting dynamos are capable of low loss contactless pumping large dc currents into superconducting magnet coils. We present a mathematical model of such devices and use it for computing the two-dimensional loop currents and electric fields induced by a dynamo rotor-mounted permanent magnet in a thin superconducting stator strip. Two numerical methods for thin film superconductivity problems, the mixed finite element method and the fast Fourier transform based method, are employed and compared. We find voltages generated in the stator under open circuit conditions and losses in the stator. It is shown that if the length of a permanent magnet is comparable to or smaller than the strip width, the one-dimensional model employed in the previous works can be inaccurate.

*Index Terms*— superconducting dynamo, coated conductor, numerical solution, nonconforming finite elements, fast Fourier transform.

## I. Introduction

THE high temperature superconducting (HTS) dynamo proposed in [1] and attracted much recent interest [2]-[9], is a kind of flux pumps (see [10] and the references therein) for low loss contactless loading, unloading, and current decay compensation in HTS magnets. The physical origin of a nonzero averaged voltage generated in the HTS dynamo stator made of a coated conductor strip has been clarified in several recent works (see [2], [6-8]): the voltage appears due to the induced overcritical currents and nonlinear resistivity of a superconducting material. Several prototypes have been successfully tested [1-10].

Two types of mathematical models have been employed for numerical simulations. Models of the first type use a simplified lumped circuit-type description of the HTS dynamo [3]. In models of the second type the stator is presented as an infinite thin HTS strip subjected to the time-varying non-uniform field of a long permanent magnet (PM) attached to the dynamo rotor [6]-[8], [11]. Although less simplified, these models assume that the eddy current density and electric field are the same in each cross section of the stator. In reality, the eddy currents and electric fields induced by the moving PM of a finite length are strong only in a stator part close to the rotor and, to estimate an induced voltage, an effective "active length" of the stator needs to be introduced [12]. Typically, the effective length is taken equal to the length of the rectangular PM [6], [8], [11]. In the infinitely thin strip approximation, fully justified for the coated conductor stator, the sheet current density in such models depends on only one spacial variable. A comparison of various numerical methods derived for this one-dimensional (1D) mathematical model was presented in [12], [13], where the problem was solved for either an open circuit condition or for a given transport current. We would like to emphasize here that we call this problem 1D although some of numerical schemes in [12] employ two-dimensional (2D) formulations in the plane orthogonal to the stator strip. Other, more efficient methods in [12], [13], are based on the formulations fully confined to the 1D cross section of the thin strip.

Unlike the previous works, here we solve the 2D superconducting film eddy current problem in the film plane and compute the loop currents induced in the stator strip by the moving PM of a finite length. We calculate also the electric field and use it to estimate the main characteristic of a dynamo, the averaged voltage generated in the stator under the open circuit condition. Our results are compared to those obtained using the 1D model.

## II. 2D Model of an HTS Dynamo

Let a uniformly magnetized PM having a rectangular prism shape be attached to the HTS dynamo rotor rotating counter clockwise around the $x$-axis, see Fig. 1.

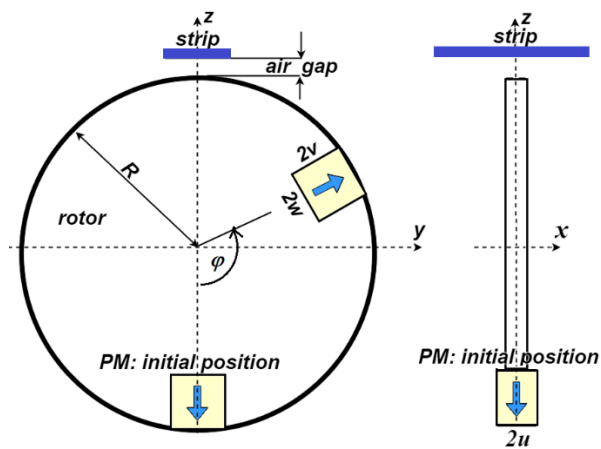

Fig.1 Schematic drawing of the HTS dynamo: projections onto the yz (left) and xz (right) planes; the thin stator strip is $\{|x|\leq b, |y|\leq a, z = z_s\}$.

Manuscript receipt and acceptance dates will be inserted here.
L. Prigozhin is with Blaustein Institutes for Desert Research, Ben-Gurion University of the Negev, Sde Boqer Campus 84990, Israel
(e-mail: leonid@bgu.ac.il).

V. Sokolovsky is with Physics Department, Ben-Gurion University of the Negev, Beer-Sheva 84105, Israel (e-mail: sokolovv@bgu.ac.il)

In the initial position the PM edges parallel to the $x, y, z$ axes are, respectively, $2u, 2w, 2v$ and the PM face plane coordinates are $x_1 = -u, x_2 = u$, $y_1 = -w, y_2 = w$, and $z_1 = -R$, $z_2 = -R + 2v$, where $R$ is the rotor radius. At a point $\boldsymbol{p} = (x, y, z)^T$ outside the magnet the magnetic field $\boldsymbol{h}_0 = (h_{0x}, h_{0y}, h_{0z})^T$ has the components, see [14],

$$h_{0x}(\boldsymbol{p}) = \frac{B_r}{4\pi\mu_0} \sum_{k=1}^{2}\sum_{m=1}^{2} (-1)^{k+m} \ln[\Phi(\boldsymbol{p}, x_m, y_1, y_2, z_k)],$$

$$h_{0y}(\boldsymbol{p}) = \frac{B_r}{4\pi\mu_0} \sum_{k=1}^{2}\sum_{m=1}^{2} (-1)^{k+m} \ln[\Psi(\boldsymbol{p}, x_1, x_2, y_m, z_k)], \quad (1)$$

$$h_{0z}(\boldsymbol{p}) = \frac{B_r}{4\pi\mu_0} \sum_{k=1}^{2}\sum_{n=1}^{2}\sum_{m=1}^{2} (-1)^{k+m+n} \arctan[\Gamma(\boldsymbol{p}, x_n, y_m, z_k)],$$

where $\mu_0$ is the permeability of vacuum, $B_r$ is the PM remanent flux density,

$$\Phi(\boldsymbol{p}, x_m, y_1, y_2, z_k) = \frac{y - y_1 + \left[(x-x_m)^2 + (y-y_1)^2 + (z-z_k)^2\right]^{1/2}}{y - y_2 + \left[(x-x_m)^2 + (y-y_2)^2 + (z-z_k)^2\right]^{1/2}},$$

$$\Psi(\boldsymbol{p}, x_1, x_2, y_m, z_k) = \frac{x - x_1 + \left[(x-x_1)^2 + (y-y_m)^2 + (z-z_k)^2\right]^{1/2}}{x - x_2 + \left[(x-x_2)^2 + (y-y_m)^2 + (z-z_k)^2\right]^{1/2}},$$

$$\Gamma(\boldsymbol{p}, x_n, y_m, z_k) = \frac{(x-x_n)(y-y_m)}{(z-z_k)\left[(x-x_n)^2 + (y-y_m)^2 + (z-z_k)^2\right]^{1/2}}.$$

After rotation of the rotor by an angle $\varphi$ the PM field at the point $\boldsymbol{p}$ becomes

$$\boldsymbol{h}_{PM}(\boldsymbol{p}) = A(\varphi)\boldsymbol{h}_0(A(-\varphi)\boldsymbol{p}), \quad (2)$$

where

$$A(\varphi) = \begin{pmatrix} 1 & 0 & 0 \\ 0 & \cos(\varphi) & \sin(\varphi) \\ 0 & -\sin(\varphi) & \cos(\varphi) \end{pmatrix}$$

is the matrix of rotation.

We neglect the stator strip thickness and present the strip as the 2D domain

$$S = \{|x| \le b, |y| \le a, z = z_s\},$$

where $2a$ and $2b$ are the strip width and length, respectively, $z_s = R + d$, and $d$ is the air gap. We assume the superconducting film is characterized by the power current-voltage relation between the component of the electric field tangential to the film, $\boldsymbol{e}$, and the sheet current density $\boldsymbol{j}$:

$$\boldsymbol{e} = e_0(|\boldsymbol{j}|/j_c)^{n-1}\boldsymbol{j}/j_c. \quad (3)$$

Here $e_0 = 10^{-4}$ Vm$^{-1}$, the power $n$ is a given constant, $j_c$ is the critical sheet current density. By the Faraday law

$$\mu_0 \dot{h}_z = -\nabla \times \boldsymbol{e}, \quad (4)$$

where $\nabla \times \boldsymbol{e} = \partial_x e_y - \partial_y e_x$, $\dot{f}$ means $\partial_t f$, and $h_z$ is the normal to the film component of the total magnetic field. Using the Biot-Savart law we obtain for the points in the film plane

$$h_z = h_{PM,z} + \nabla \times \int_S G(\boldsymbol{r}-\boldsymbol{r}')\boldsymbol{j}(\boldsymbol{r}')\,d\boldsymbol{r}'. \quad (5)$$

Here $\boldsymbol{r} = (x, y)$ and $G(\boldsymbol{r}) = (4\pi|\boldsymbol{r}|)^{-1}$ is the Green function. In addition, the sheet current density should satisfy the conditions

$$\nabla \cdot \boldsymbol{j} = 0 \text{ in } S, \quad \boldsymbol{j} \cdot \boldsymbol{v}|_{\partial S} = 0, \quad (6)$$

where $\boldsymbol{v}$ is a normal to the boundary $\partial S$ of $S$. Equations (1)-(6) constitute the 2D HTS dynamo model considered in this work.

We note also that due to (6) there exists the stream function $g$ such that

$$\boldsymbol{j} = \bar{\nabla} \times g \quad (7)$$

in $S$, i.e. $j_x = \partial_y g$, $j_y = -\partial_x g$, and $g = 0$ on $\partial S$. The critical sheet current density in (3) can depend on the magnetic field but, for simplicity, we will assume a constant $j_c$. Our aim is to solve the problem (1)-(6) numerically and to compute the averaged voltage

$$V(t) = \frac{1}{2a}\int_S e_x(t,\boldsymbol{r})\,d\boldsymbol{r}. \quad (8)$$

### III. Two Numerical Methods

Numerical solution of 2D thin film problems in superconductivity can be obtained for a formulation in terms of the stream (magnetization) function using the finite Fourier series [15] or a finite element method, see e.g. [16]. The sheet current density is then determined as a 2D curl of this function and the Biot-Savart law can be used to find the magnetic field. However, determining the electric field by such approaches remains a problem. Indeed, although the magnetization function can be found with high accuracy, accuracy of the sheet current density is much lower because its determination involves numerical differentiation. Significant further multiplication of error occurs if a typical of type-II superconductors highly nonlinear current-voltage relation is used then to calculate the electric field.

To circumvent this difficulty a mixed variational formulation for thin film problems, written in terms of two variables, the electric field and the magnetization function, has been derived [17]; numerical approximation in that work employed the implicit finite difference approximation in time, piecewise constant elements for the magnetization function, and the Raviart-Thomas elements of the lowest order for the electric field. A simpler numerical scheme proposed in [18] is also based on the mixed formulation but uses the nonconforming piecewise linear elements for the magnetization function and piecewise constant vectorial elements for the electric field.

Completely different numerical method, using the method of lines for integration in time and the fast Fourier transform (FFT) based approximation in space, has been proposed for thin film magnetization problems in [19], [20] and modified in [21]. This method is written for a formulation in terms of the magnetization function alone and fails, in general, to accurately compute the electric field [21]. Nevertheless, as was noticed in [21], the electric field can be found with sufficient accuracy in problems, where the film boundary is well aligned with the regular 2D mesh.

The latter is true for a rectangular film, the dynamo stator, considered in our work. Using the 2D HTS dynamo model we compare these two numerical methods, the mixed finite element method [18] and the modified FFT-based method [21].

### A. The Mixed Method

Let us replace the curl-conforming electric field $e$ by an auxiliary divergence-conforming variable $q$ setting $q_x = e_y, q_y = -e_x$. Then $\nabla \times e = -\nabla \cdot q$ and, using (4), (5) and (7) we obtain

$$-\mu_0^{-1}\nabla \cdot q = \dot{h}_{PM,z} + \nabla \times \int_S G(r-r')\bar{\nabla}' \times \dot{g}(t,r')\,dr' \quad (9)$$

Multiplying this equation by a sufficiently smooth function $\psi(r)$ satisfying $\psi|_{\partial S} = 0$, integrating and using Green's theorem, we arrive at the variational equation

$$\mathcal{A}(\dot{g},\psi) - \mu_0^{-1}(q,\nabla \psi) = -(\dot{h}_{PM,z},\psi), \quad (10)$$

where $(f,g) = \int_S f \cdot g\,dr$ is the scalar product of two vector functions, similarly for the scalar case, and the bilinear form

$$\mathcal{A}(f,\psi) = (4\pi)^{-1} \iint_{S\,S} \frac{\nabla f(r) \cdot \nabla' \psi(r')}{|r-r'|}\,dr\,dr'.$$

Inverting equation (3) we obtain

$$j = j_c(|e|/e_0)^{\kappa-1}e/e_0$$

with $\kappa = 1/n$. Noting that $|e| = |q|$ and using (7) we find, after simple transformations,

$$\nabla g = -j_c(|q|/e_0)^{\kappa-1}q/e_0. \quad (11)$$

The simplest approximation of the mixed variational formulation (10)-(11) would employ continuous piecewise linear elements for the magnetization function $g$ and piecewise constant vectorial elements for the rotated electric field $q$. However, such combination yields a 'mosaic' structure of the electric field which makes the calculated field useless. The situation is typical for the critical state problems; see, e.g., figure 1 in [22]. Our stable approximation, applied to the time discretized version of (10),

$$\mathcal{A}(g^m - g^{m-1},\psi) - \tau^m \mu_0^{-1}\left(\frac{q^m + q^{m-1}}{2}, \nabla \psi\right) =$$
$$-(h_{PM,z}^m - h_{PM,z}^{m-1}, \psi),$$

uses the nonconforming piecewise linear elements for $g$ (linear in each triangular element and continuous in the middle of its edges) with the piecewise constant vectorial elements for $q$. Here $m$ is the time layer number and $\tau^m$ is the time step. We note that this scheme has the second order approximation in time; that is the only difference with [18], where the fully implicit first order approximation was used. The finite element approximation of the bilinear form $\mathcal{A}$ needs accurate compution of the double surface integrals

$$K_{k,l} = \iint_{s_k\,s_l} \frac{dr\,dr'}{|r-r'|} \quad (12)$$

for all pairs of trangular elements $s_k$, $s_j$. Some of these integrals are singular; as in [18], we followed [23] for their calculation. The nonlinearity in (11) was, on each time layer, resolved iteratively by replacing $|q|^{\kappa-1}q$ with $|q^{(i-1)}|^{\kappa-1}q^{(i-1)} + (|q^{(i-1)}|_\varepsilon)^{\kappa-1}(q^{(i)}-q^{(i-1)})$, where $i$ is the iteration number and $|q|_\varepsilon = \sqrt{|q|^2 + (e_o\varepsilon)^2}$ with a small $\varepsilon > 0$; we used $\varepsilon = 10^{-10}$ and an over-relaxation to accellerate convergence. We refer to [18] for further details of the numerical algorithm.

### B. The FFT-based Method

Extending the sheet current density $j$ and magnetization function by zero outside the film domain $S$ in the plane $z = z_s$ one can write (5) by means of 2D convolutions as $h_z = h_{PM,z} + (\partial_x G) * j_y - (\partial_y G) * j_x$ or, in terms of the magnetization function, $h_z = h_{PM,z} - (\partial_x G)*(\partial_x g) - (\partial_y G)*(\partial_y g)$. Taking Fourier transform and noting that $\mathcal{F}[G] = (2|k|)^{-1}$, where $k = (k_x, k_y)$, we arrive at $\mathcal{F}[h_z - h_{PM,z}] = (|k|/2)\mathcal{F}[g]$. Hence

$$g = \mathcal{F}^{-1}\left[\frac{2}{|k|}\mathcal{F}[h_z - h_{PM,z}]\right] - C, \quad (13)$$

where for $k = 0$ we replace $2/|k|$ by zero and introduce a time-dependent constant $C(t)$ determining the $\mathcal{F}[g]$ value for $k = 0$. Since $g$ should be zero in $S_{out} = \mathbb{R}^2 \setminus S$ we fix $C(t)$ using the condition $\int_{S_{out}} g\,dr = 0$. Differentiating with respect to time we obtain a differential equation for $g$,

$$\dot{g} = \mathcal{F}^{-1}\left[\frac{2}{|\boldsymbol{k}|}\mathcal{F}\left[\dot{h}_z - \dot{h}_{\text{PM},z}\right]\right] - \dot{C}, \qquad (14)$$

where the unknown shift $\dot{C}$ is determined by the integral condition $\int_{S_{\text{out}}} \dot{g}\,d\boldsymbol{r} = 0$. If $g$ for time $t$ is known, the sheet current density is determined using (7) and the electric field inside the strip $S$ by (3). The Faraday law (4) determines then $\dot{h}_z$ in $S$.

To integrate (14) in time one needs to find $\dot{h}_z$ also in $S_{\text{out}}$. The problem can be formulated as follows: find $\dot{h}_z$ in $S_{\text{out}}$ such that equation (14) holds with $\dot{g}=0$ in $S_{\text{out}}$.

Initial approximation to this problem solution is based on assignment to $S_{\text{out}}$ a high constant resistivity $\rho_{\text{out}}$ to suppress the appearance of stray currents in this domain. Using such an approximation alone is known to be not fully satisfactory: if the artificial resistivity $\rho_{\text{out}}$ is not sufficiently large, the stray current appears, if it is high the problem becomes stiff and its integration in time slows down. Hence it is better to choose a moderately high $\rho_{\text{out}}$ and use iterations to subtract from $\dot{h}_z$ in $S_{\text{out}}$ its part related to $\partial_t \boldsymbol{j}|_{S_{\text{out}}} = \overline{\nabla} \times \dot{g}|_{S_{\text{out}}}$. The role of $\rho_{\text{out}}$ in this scheme is twofold: it provides a good initial approximation for further iterations and, as showed our numerical experiments, is also necessary to suppress the stray current in a very narrow layer near the film boundary $\partial S$.

For practical implementation of this scheme a regular $N_x \times N_y$ grid should be defined in a rectangular computation domain containing and several times larger than the strip domain $S$. Values of all variables are defined at the grid nodes; the continuous Fourier transform and its inverse are replaced by their discrete analogs on this grid and computed using the FFT algorithm. The 2D curl operators $\overline{\nabla} \times \boldsymbol{f}$ and $\nabla \times \boldsymbol{f}$ are computed in the Fourier space with the Gaussian smoothing to suppress high frequency oscillations. As in [21], the parameter of smoothing in our simulations was equal to the length of the grid cell diagonal. Weaker smoothing does not change the results much but significantly increases the computation time. The same Gaussian smoothing was applied on iterations involving $\dot{h}_z$ calculations. Finally, the Matlab ordinary differential equation (ODE) solver *ode23* with an automatic time step control and default parameter values was used to integrate the spacially discretized problem (14) in time. More details on our algorithm can be found in [21]; see also [24], where this method was extended to the film stack problems.

All numerical simulations in our work were done in Matlab R2020a on a PC with the Intel (R) Core (TM) i7-4770 CPU@ 3,40 Hz and 16 GB RAM.

## IV. SIMULATION RESULTS

We solved the 2D HTS dynamo problem with the same parameters as in the 1D problem in [12], [13] but for the finite lengths $2u$ of the PM and $2b$ of the stator strip (Table I).

TABLE I
HTS DYNAMO BENCHMARK PARAMETERS

| | | |
|---|---|---|
| Permanent magnet (PM) | width, $2w$ | 6 mm |
| | height, $2v$ | 12 mm |
| | length, $2u$ | 12.7 mm |
| | remanent flux density, $B_r$ | 1.25 T |
| HTS stator strip | width, $2a$ | 12 mm |
| | length, $2b$ | 48 mm |
| | thickness | 1 μm |
| | critical sheet current density, $j_c$ | 23.6 A/mm |
| | power value, $n$ | 20 |
| rotor external radius, $R$ | | 35 mm |
| air gap, $d$ | | 3.7 mm |
| frequency of rotation, $f$ | | 4.25 Hz |

In the 1D model, for estimating the induced voltage an "active zone length" is introduced and assumed, somewhat arbitrary, equal to the PM length, $2u$. Our 2D model accounts for the real PM length. The considered part of the stator should be long enough to include the area of any singificant current density and electric field; our simulations (see below) showed that for this PM the chosen stator length is sufficient.

We assumed the open circuit condition and zero initial sheet current density. For the mixed method the iterations were performed till their convergence with the relative tolerance $5 \cdot 10^{-4}$ (in the $L^1$-norm). The chosen time step $\tau^m$ corresponded to the rotor turn by $0.2^\circ$ when the PM is close to the stator and was much larger otherwise.

For the FFT-based method the stator $\{|x| \leq 4a,\ |y| \leq a\}$ was placed inside the computation domain $D = \{(x,y): |x| \leq 6a,\ |y| \leq 2a\}$, the auxiliary resistivity was taken as $\rho_{\text{out}} = 3 \cdot 10^3 e_0 / j_c$, and the relative tolerance for iterations set to $2 \cdot 10^{-4}$. The integration time step was chosen automatically and the approximate values of $\dot{h}_{\text{PM},z} = 2\pi f\, \partial h_{\text{PM},z}/\partial\varphi$ at all grid nodes were interpolated from the look up table with the angle step $\Delta\varphi = 1^\circ$. As in [21], this table was derived from a table of the grid $h_{\text{PM},z}(\varphi, x, y, z_s)$ values using numerical differentiation in the Fourier space.

To compare the two methods, for each of them we performed simulation of two cycles of rotor revolution using several meshes. Meshes for the mixed method where nonuniform (refined in the subdomain $\{(x,y): |x| \leq 2.4a,\ |y| \leq a\}$ containing the significant electric field area).

The computation times per cycle and the mean voltage, which is the most important characteristic of a superconducting to $580^0$ during the second cycle and similar rotor positions for all other cycles. This means that up to six equidistantly mounted PMs act independantly and the voltages induced by each of them can be simply summed up. For a larger number of PMs the resulting time-averaged voltage is expected to be less than the corresponding sum; this observation agrees with the experiment [9].

TABLE II
NUMERICAL SIMULATION RESULTS

| **MIXED METHOD** | | |
|---|---|---|
| N OF FINITE ELEMENTS | COMPUTATION TIME PER CYCLE, MINUTES | MEAN VOLTAGE FOR THE 2-ND CYCLE, $\mu V$ |
| 800 | 3.2 | -9.03 |
| 2180 | 39 | -9.25 |
| 3286 | 115 | -9.29 |
| **FFT-BASED METHOD** | | |
| MESH IN THE COMPUTATION DOMAIN | COMPUTATION TIME PER CYCLE, MINUTES | MEAN VOLTAGE FOR THE 2-ND CYCLE, $\mu V$ |
| 648x216 | 27 | -9.29 |
| 1024x384 | 181 | -9.18 |
| 1536x512 | 757 | -9.15 |

Simulated distributions of the electric field component $e_x$ and sheet current density $j$ for three characteristic positions of the rotor during its second turn are presented in Figures 3 and 4, respectively.

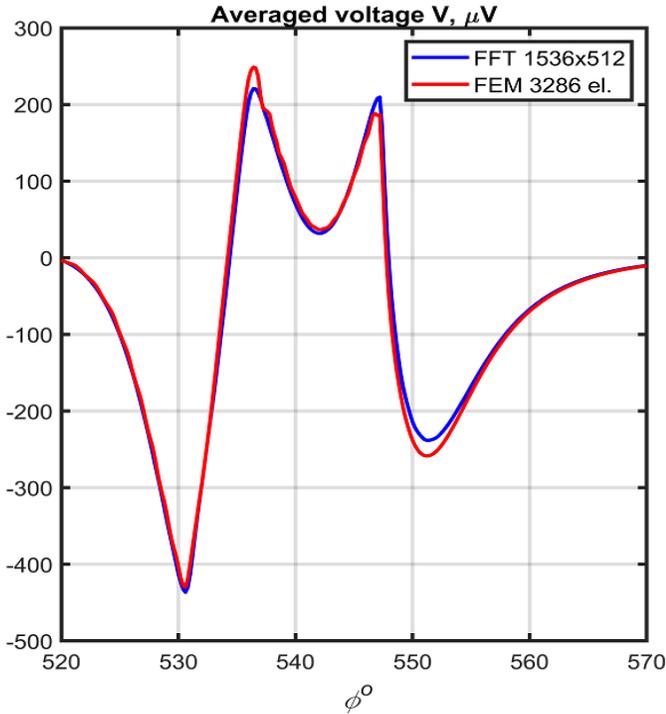

Fig. 2. Averaged voltage curves computed using the mixed finite element method (red) and the FFT-based method (blue) for the rotor rotation angles of the second turn.

dynamo, are presented in Table II. As in the previous works [12], [13], the curves of the averaged voltage $V$ (Fig. 2) are presented for the second rotor revolution cycle. The solutions obtained by the two methods are close. Using these results, it is difficult to determine which method is more accurate. The mixed method turned out to be faster in our simulations. Contrary to the FFT-based method, the mixed method needs neither selection of an artificial resistivity $\rho_{out}$, nor smoothing. The mixed method also needs no extended computation domain and allows using a nonuniform mesh. For finer grids the FFT based method quickly leads to a stiff ODE system and the computation time increases (Table II). On the other hand, implementation of the FFT-based method can be simpler because no computing the singular double-surface integrals (12) is needed.

The calculated averaged voltages (see Fig. 2) are significant only for the rotor positions corresponding to angles from $520^0$

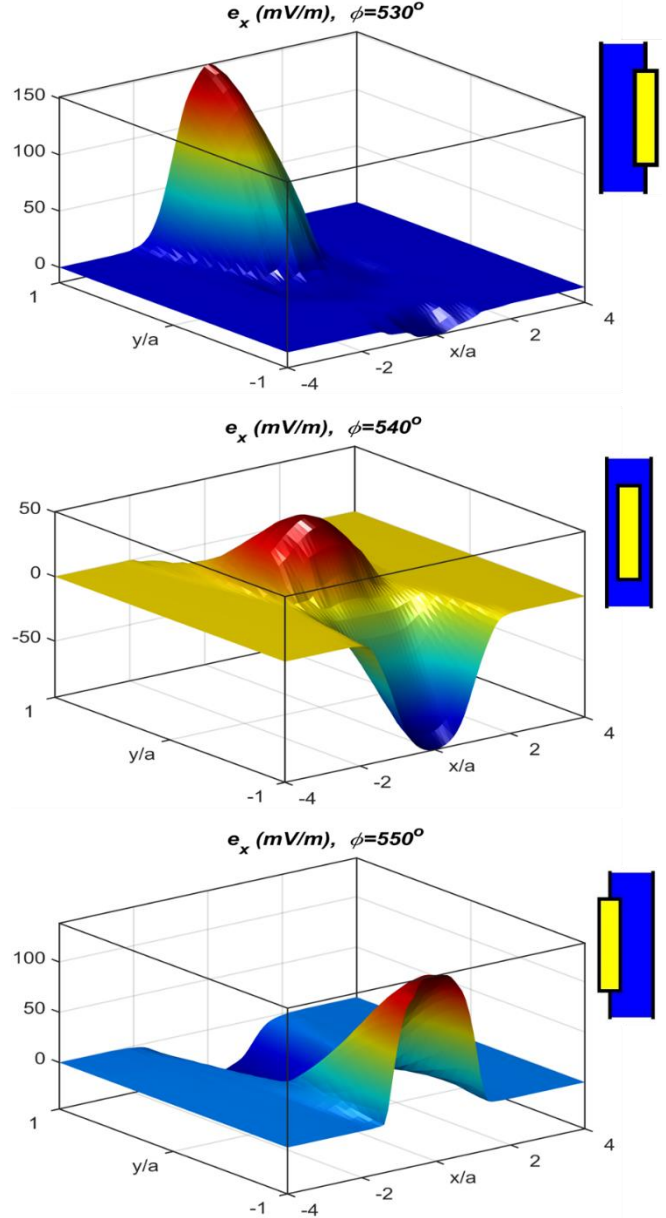

Fig. 3. Electric field component $e_x$ for three rotor positions. The insets on the right show the corresponding PM positions. Simulation using the mixed method.

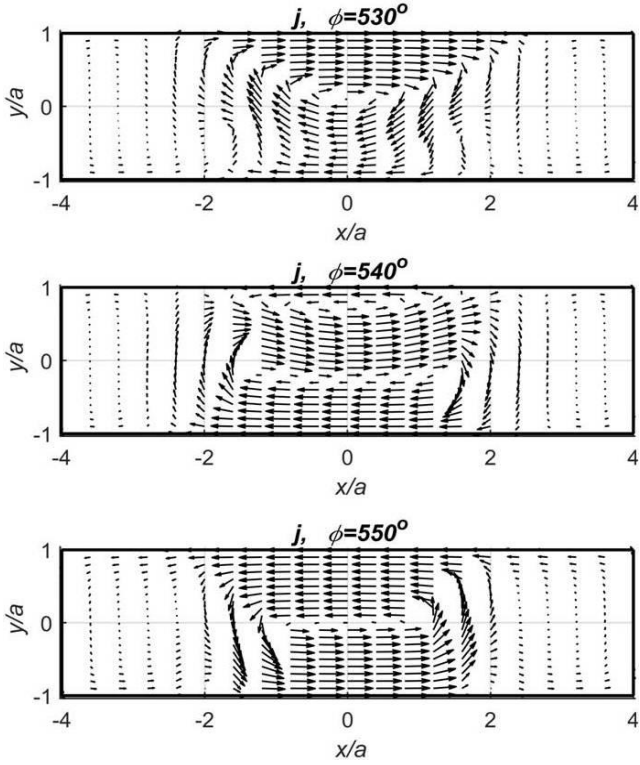

Fig. 4. Sheet current density for the three rotor positions as in Fig. 3; simulation using the mixed method.

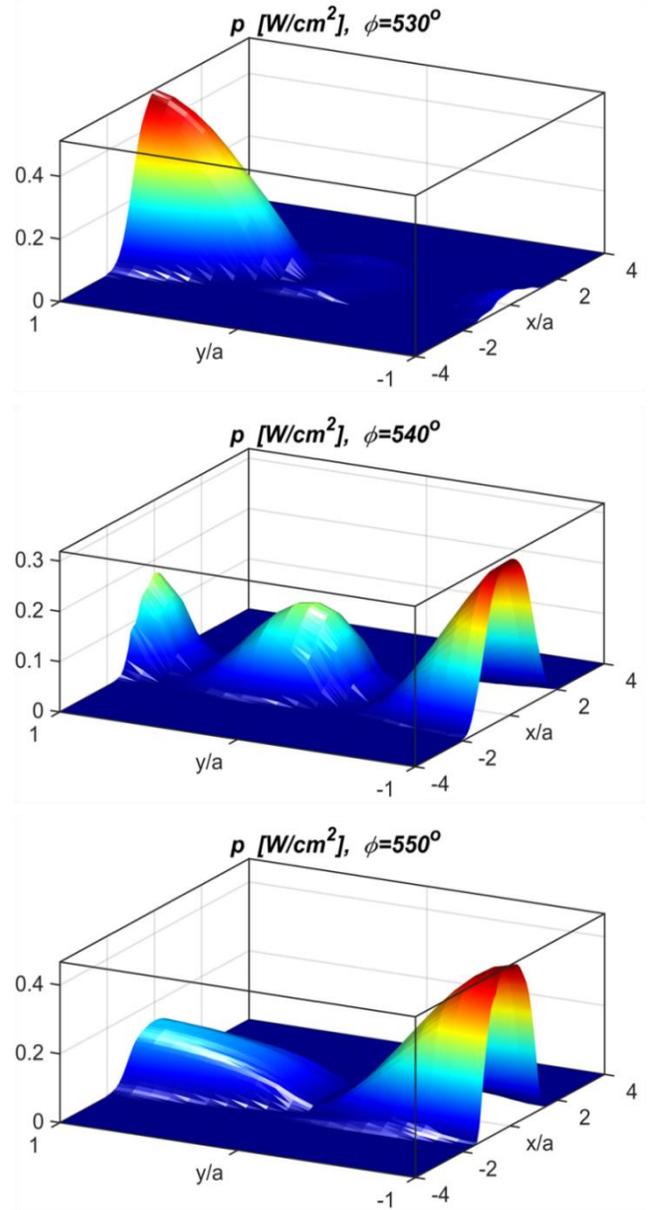

Fig. 5. Loss power density in the stator strip for three rotor positions; the mixed method.

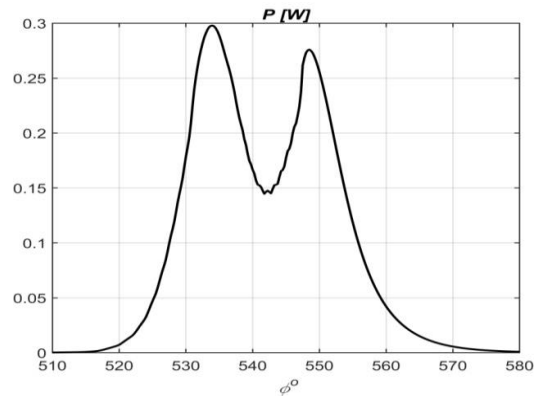

Fig. 6. Total loss power in the stator during the second cycle; the mixed method.

Clearly, no noticeble electric field or current appears in the stator strip outside the area about twice longer than the length of the PM; hence, the chosen length of the simulated stator part is fully sufficient for these simulations. The component $e_y$ in our simulation remained less than 15 mV/m. A movie showing the simulated evolution of the $e_x$ distribution during the first two rotor cycles is available as a supplementary material.

Local loss can cause thermal instabilities, appearance and propagation of normal zones (see [25] and the references therein). Hence, it is desirable to know also the instantaneous loss power density $p(t,\boldsymbol{r}) = \boldsymbol{e} \cdot \boldsymbol{j}$; its distributions corresponding to three rotor positions are presented in Fig. 5. We also computed the loss power $P(t) = \int_S \boldsymbol{e} \cdot \boldsymbol{j} \, \mathrm{d}\boldsymbol{r}$ (Fig. 6). The total loss during the second cycle was estimated as $W = 4.3\,\mathrm{mJ}$; the time-averaged loss power is, therefore, $fW = 18.3\,\mathrm{mW}$.

We note, that the loss power is approximately proportional to $f^2$ and to the number of PMs attached to the rotor. The HTS dynamo constructed in [27] contains 6 magnets and the rotor frequency can achieve 41.6 Hz. Hence, the loss power in the stator can be estimated as 11 W and can increase further in the presence of transport current; intensive cooling is required.

Using the mixed method for the 2D dynamo model and the Chebyshev-polynomial-based method [13] for the 1D model, we compared the simulated voltage curves for these two models (Fig. 7) and found that the difference is significant if a magnet length does not exceed the strip width (Fig. 7, top). The magnet in the benchmark problem [12],[13], assumed also in our simulations above, is slightly longer than the strip width and the two curves become closer to each other (Fig. 7, middle). The mean voltage, computed for the second cycle using the 1D model, is $-10.0\,\mu V$, which is not too different from the 2D model values in Table II. As could be expected, the longer is the PM the more accurate the 1D model becomes (Fig. 7, bottom). However, even for a long PM the 2D model provides a much more detailed solution.

This paper was already prepared for submission when there appeared a preprint with a similar HTS dynamo model by Ghabeli, Pardo and Kapolka, "3D modeling of a superconducting dynamo-type flux pump" [27]. Using regular finite element meshes containing only one layer of 3D elements, these authors adapted to thin films their numerical scheme for 3D superconductivity problems. Although the model is called 3D, this term better describes the employed numerical method than the problem itself; using a one-layer discretization is similar to considering a 2D problem. Presented computation times indicate that our mixed method for thin superconducting films is, probably, more efficient. On the other hand, Ghabeli *et al*. performed simulations also with a field dependent critical current density.

## V. Conclusion

Two numerical methods, the mixed finite element method [18] and the FFT-based method [19], [21] have been employed for solving the 2D HTS dynamo problem; the simulation results obtained are close. For simplicity, we assumed the stator is characterized by the power current-voltage relation with a constant critical sheet current density. Both methods can be extended to problems with a field-dependent relation (for the mixed method this was realized in [28]). The mixed method is, possibly, more efficient: it was somewhat faster than the FFT method, there was no need to consider any region outside a superconductor, to introduce an artificial resistivity, and to apply smoothing. Unlike the FFT-based method, it is able to compute the electric field also in non-rectangular films [17,18].

The 1D model of an HTS dynamo [6], [12], [13] is able to produce qualitatively correct curves of the averaged voltage generated in the dynamo stator under the open circuit condition. However, this model assumes the PM and HTS strip are infinitely long, so the current distribution is the same in each cross section of the stator strip, and uses an effective length of the active zone of the stator. The length is not known *a priory* and, although our 2D simulations indicate that the usual practice of setting this length equal to the PM length is reasonable, this is hardly garanteed for all cases. If the PM is not sufficiently long, the 1D model leads to a significant error in computing the voltage generated by a dynamo (Fig. 7).

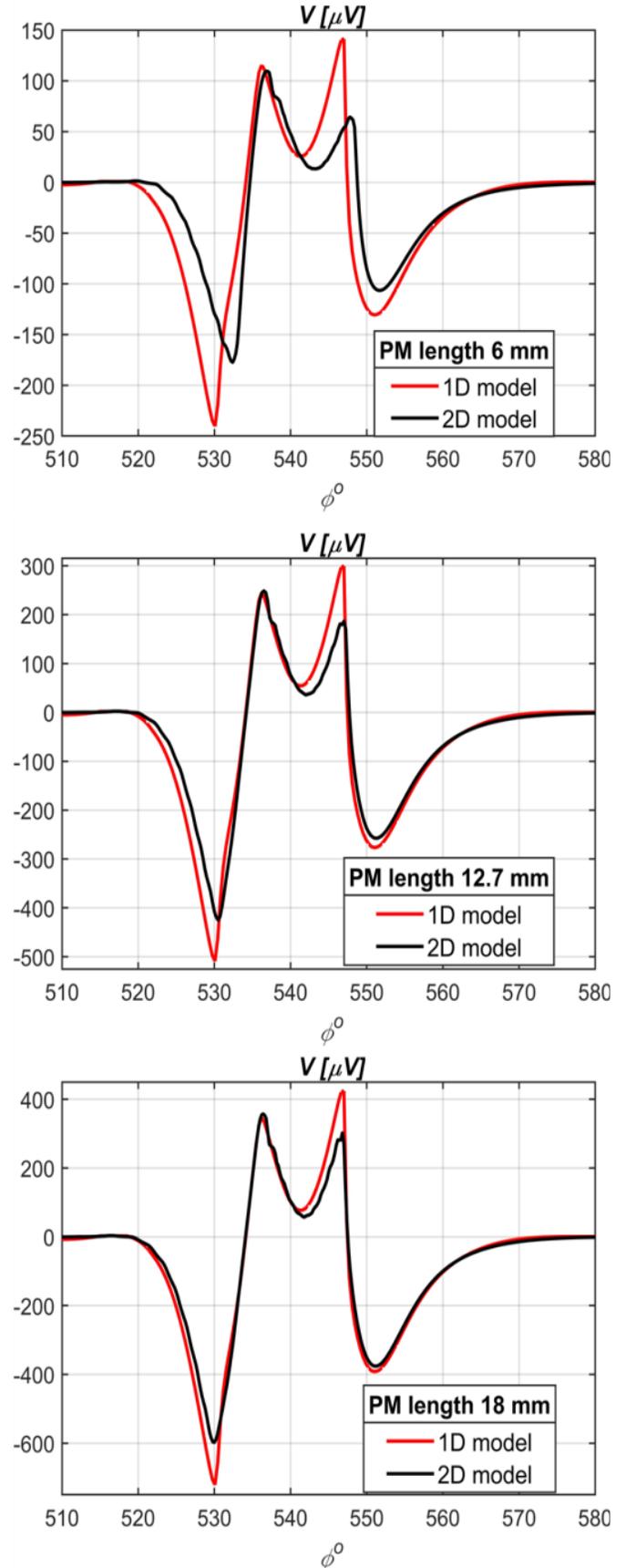

Fig. 7. Averaged voltage curves for PMs of different lengths; simulation results for 1D and 2D models.

The more realistic 2D model presented in our work avoids the main simplification of the 1D model and is more accurate. Taking into account both components of the current density to describe the induced eddy current loops, this model can provide better estimates for losses in the stator strip. The model is easily adapted to the multi-PM case and can be used to determine whether the much simpler 1D model, allowing one to simulate thousands of rotor revolutions in a closed-contour dynamo simulation, could be reliable. Such models can help to optimize the HTS dynamo design and provide a useful guidance also for constructing flux pumps of other types, e.g., where a coated conductor is subjected to a traveling wave external magnetic field.

## SUPPLEMENTED MATERIALS

A movie showing the simulated evolving distribution of the electric field is available as
https://www.math.bgu.ac.il/~leonid/FEM_Ex_movie.avi